\documentclass[sigconf,natbib=true]{acmart}
\usepackage{enumitem}
\usepackage{natbib}
\usepackage[table]{xcolor}
\usepackage{multirow}
\usepackage{tikz}
\usepackage{pgfplots}

\copyrightyear{2026}
\acmYear{2026}
\setcopyright{cc}
\setcctype{by}
\acmConference[ICTIR '26]{Proceedings of the 2026 International ACM SIGIR Conference on Innovative Concepts and Theories in Information Retrieval (ICTIR)}{July 25, 2026}{Melbourne, VIC, Australia}
\acmBooktitle{Proceedings of the 2026 International ACM SIGIR Conference on Innovative Concepts and Theories in Information Retrieval (ICTIR) (ICTIR '26), July 25, 2026, Melbourne, VIC, Australia}
\acmDOI{10.1145/3805713.3820411}
\acmISBN{979-8-4007-2600-2/2026/07}

\begin{document}

\title[Entity Labels Are Not Entity Signals]{Entity Labels Are Not Entity Signals: A Framework for Observable Relevance in Document Re-Ranking}
\author{Utshab Kumar Ghosh}
\affiliation{%
  \department{Department of Computer Science}
  \institution{Missouri University of Science and Technology}
  \city{Rolla}
  \state{MO}
  \country{USA}
}
\orcid{0000-0003-3096-6909}
\email{u.ghosh@mst.edu}

\author{Shubham Chatterjee}
\affiliation{%
  \department{Department of Computer Science}
  \institution{Missouri University of Science and Technology}
  \city{Rolla}
  \state{MO}
  \country{USA}
}
\orcid{0000-0002-6729-1346}
\email{shubham.chatterjee@mst.edu}

\renewcommand{\shortauthors}{Ghosh and Chatterjee}

\begin{abstract}
Entity-aware document retrieval uses query-associated entities as ranking
signals, assuming that semantically relevant entities are also useful
retrieval signals. We show this assumption is insufficient---and explain
why. Unlike terms, which are ground-truth observations, entity links are
hypotheses produced by an imperfect linker: an entity can be topically
central yet provide no discriminative signal if the linker fires
indiscriminately across relevant and non-relevant documents. We formalize
this as a distinction between \emph{Conceptual Entity Relevance}
(CER)---whether an entity is topically related to a query---and
\emph{Observable Entity Relevance} (OER)---whether its observed presence
in a collection discriminates relevant from non-relevant documents. Across
four collections and annotation sources including human entity judgments,
CER and OER exhibit near-chance agreement ($\kappa \approx 0$), while OER
operationalizations agree substantially ($\kappa \approx 0.5$), confirming
CER as the systematic outlier. CER-based supervision selects topically
plausible but weakly discriminative entities, pruning fewer than 4\% of
non-relevant documents on some collections. Aligning supervision with OER
improves non-relevant pruning by up to 10$\times$ and open-world MAP by
0.051 over BM25. Our findings motivate a shift from conceptual to
observable notions of entity relevance in entity-aware retrieval.
\end{abstract}



\maketitle

\section{Introduction}
\label{sec:intro}

Entity-aware retrieval uses query-associated entities as signals for document
ranking. This raises a central question: \emph{which entities should influence
ranking?} The wrong entities may produce misaligned signals, so the key issue is
not whether entities can help retrieval, but what makes an entity the
\emph{right} signal. Existing
work~\cite{xiong2017word,liu-etal-2018-entity,liu2015latent,dalton2014entity,xiong2015esdrank,chatterjee2025qder,chatterjee2024dreq,tran2022dense}
has largely answered this question in terms of topicality: an entity is treated
as useful if it is \emph{about} the query---the notion naturally captured by
human annotation and LLM judgments~\cite{saliminabi2025llm}. We call this
\emph{Conceptual Entity Relevance} (CER). This assumption is reasonable in
principle: if a document discusses the query topic, it should mention topically
related entities. The problem is that entity-aware retrieval does not operate
directly on topics---it operates on \emph{linked entities}, and linking is
noisy.

This gap partly resembles a well-known phenomenon in term-based retrieval:
semantically related expansion terms are not necessarily useful retrieval terms,
and embedding-nearest terms may be beneficial, neutral, or harmful depending on
their effect on retrieval
effectiveness~\cite{voorhees1994query,carpineto2012survey,roy2016using,imani2019deep}.
Related work on locally trained word embeddings further shows that global
semantic similarity can be too coarse for retrieval, and that query- or
collection-specific evidence can yield better expansion
terms~\cite{diaz-etal-2016-query}.

For entities, however, the problem is sharper than it is for terms. A term
occurrence is directly observed: a term either appears in a document or it does
not, and its presence is a ground-truth fact. Entity occurrences must be
inferred by a linker, and this inference introduces a layer of uncertainty that
terms never face. The linker may miss a surface form, map a mention to the wrong
entity, or fire on an incidental reference. As a result, an entity can be
topically central yet non-discriminative if the linker fires indiscriminately
across relevant and non-relevant documents; conversely, a peripheral entity can
be a strong signal if its observed mentions concentrate in relevant documents.
CER asks whether an entity is about the query, but ignores the observation layer
entirely---how the linker fires, and whether those firings concentrate in
relevant documents. This gap is not a matter of label noise or model quality.
It is structural.

We therefore distinguish CER from \emph{Observable Entity Relevance} (OER).
OER asks whether observing an entity, as linked in a specific collection,
helps distinguish relevant from non-relevant documents. OER depends on how the
entity is actually observed in the retrieval setting: whether it is linked, how
often, and whether those links concentrate in relevant documents. The CER--OER
gap is therefore not merely label noise or model weakness; it is a mismatch
between topical plausibility and discriminative retrieval evidence.

Our central claim is that entity-aware retrieval has relied on CER as a proxy
for downstream retrieval utility, even though effective document ranking depends
on discriminative evidence rather than topical plausibility. To study this
mismatch, we introduce a diagnostic framework for analyzing the discriminative
utility of entity signals, and show that the CER--OER gap is large, structured,
and general across collections and annotation sources.

This label-level mismatch has direct downstream consequences. CER-based
approaches emphasize topically plausible but weakly discriminative signals.
These signals can appear effective under \emph{closed-world} evaluation, where
ranking is restricted to documents containing at least one selected entity.
However, they transfer poorly to \emph{open-world} settings, where signals must
operate over the full candidate set. This reflects a mismatch in supervision and
evaluation, rather than a limitation of modeling. We use OER as a diagnostic and
show that approximations to it yield stronger entity signals and improved
open-world retrieval.

\smallskip
\noindent We make the following \textbf{contributions}\footnote{Code, data, and
prompts: \url{https://github.com/shubham526/ICTIR2026-CER-vs-OER}}:
\begin{enumerate}[leftmargin=*]

\item We distinguish CER from OER and formalize this distinction with
coverage-based diagnostics that test whether selected entities isolate relevant
documents. We show that entity-aware retrieval introduces a structural mismatch
beyond query expansion: unlike terms, entity links are hypotheses produced by an
imperfect linker rather than direct textual observations, and CER cannot account
for how those hypotheses behave in a given collection.

\item We provide the first empirical characterization of the CER--OER gap,
showing near-chance agreement ($\kappa \approx 0$) across four collections and
annotation sources, including human entity annotations. In contrast, OER
operationalizations agree substantially more strongly ($\kappa \approx 0.5$),
showing that the disagreement is structured rather than random.

\item We show that this gap has direct downstream consequences: CER-based
selectors produce weakly discriminative entity runs, prune fewer than $4\%$ of
non-relevant documents on some collections, and yield little open-world
improvement regardless of architecture or supervision source.

\item We show that aligning supervision with OER improves open-world MAP by
$0.051$ over BM25 and non-relevant pruning by up to $10\times$, with gains that
substantially exceed the variation among supervised CER-based selectors. This
indicates a qualitative difference between CER- and OER-based supervision.

\end{enumerate}

\section{Related Work}
\label{sec:related_work}

\noindent\textbf{Entity-aware retrieval models.}
Entity-aware retrieval incorporates entities as signals for document
ranking~\cite{dalton2014entity}. Existing approaches range from methods that
explicitly rank query-relevant entities~\cite{chatterjee2024dreq,chatterjee2025qder}
to architectures that integrate entities implicitly through interaction,
attention, or embedding-based
mechanisms~\cite{xiong2017word,xiong2017jointsem,liu-etal-2018-entity,xiong2015esdrank,tran2022dense,nguyen-etal-2024-dyvo}.
Despite these architectural differences, they share a common assumption:
\emph{some entities should matter more than others for downstream ranking}.
This importance is typically grounded in topical relevance to the query.

\smallskip

\noindent\textbf{Entity relevance and supervision.}
In many systems, entity importance is implicitly learned from document-level
supervision~\cite{xiong2017word,liu-etal-2018-entity,xiong2015esdrank,tran2022dense}.
Recent methods expose this decision more directly through a separate
entity-ranking stage~\cite{chatterjee2024dreq,chatterjee2025qder}, which
requires entity-level labels. Because standard IR test collections provide
document relevance judgements rather than entity relevance judgements, such
labels must be approximated from relevant documents, structural signals such
as hyperlinks~\cite{dietz2017trec}, or LLM judgments over query--entity
pairs~\cite{saliminabi2025llm}. Although these strategies differ in form,
they largely share the same assumption: entities are useful insofar as they
are topically related to the query. This corresponds to the notion of
\emph{Conceptual Entity Relevance} (CER) formalized in this paper.

\smallskip

\noindent\textbf{Relevance vs.\ discriminativeness in retrieval.}
The broader retrieval literature has long shown that topical relevance does
not always imply discriminative utility. In term-based query expansion,
semantically related terms may be beneficial, neutral, or harmful depending
on their effect on
retrieval~\cite{voorhees1994query,carpineto2012survey,roy2016using,imani2019deep}.
Global semantic similarity can also be too coarse when query- or
collection-specific evidence would yield better
expansion~\cite{diaz-etal-2016-query}. More generally, supervision proxies
such as click logs, pseudo-labels, and first-stage outputs are effective only
when they preserve the properties that matter for
ranking~\cite{deghani2017neural,karpukhin-etal-2020-dense,xiong2020ance}.
Our work identifies a specific instance of this problem in entity-aware
retrieval. The CER--OER gap is not simply reducible to the term-based case:
terms are directly observed in text, whereas entity links are hypotheses
produced by an imperfect linker. Topical relevance can therefore fail as a
proxy not only because semantic relatedness need not imply discriminative
utility, but also because the observation layer itself is unreliable.

\smallskip

\noindent\textbf{Entity pruning in prior work.}
Many entity-aware methods restrict or denoise the entity channel before
ranking. EQFE~\cite{dalton2014entity} uses high-precision linking variants,
including top-1 entity selection and NIL classification.
JointSem~\cite{xiong2017jointsem} anchors the entity pool to query surface
forms and restricts each surface form to its top-$k$ candidates by commonness.
EVA~\cite{tran2022dense} constructs entity clusters with pairwise similarity
thresholding, explicitly arguing that unrelated entities dilute useful ones.
LES~\cite{liu2015latent} selects only the top-$k$ query-related entities,
noting that using all entities is both computationally prohibitive and
harmful. Word--Entity Duet~\cite{xiong2017word} applies soft filtering via
attention to suppress noisy query-side entities. These choices partially
shield rankers from entity-linking noise. Once the entity channel is curated,
topically plausible entities can appear useful because noisy evidence has
already been removed. Under open-world conditions, this shield is absent, and
we show that CER-based supervision fails to transfer reliably.

\smallskip

\noindent\textbf{Positioning.}
We distinguish CER---whether an entity is \emph{about} the query---from
OER---whether its observed presence in a collection helps distinguish relevant
from non-relevant documents. This distinction clarifies what closed- and
open-world evaluation measure. Closed-world evaluation asks whether a ranker
can exploit entity evidence once that evidence is available. Open-world
evaluation asks whether the same entity signals remain useful over the full
candidate set. Our paper diagnoses this mismatch and studies its consequences
for entity-aware retrieval design and evaluation.

\section{Framework}
\label{sec:framework}

We now formalize the distinction introduced above. This section defines the
retrieval setting, CER and OER, explains why they diverge, and derives the
hypotheses that structure the remainder of the paper.

\subsection{Problem Setting}
\label{subsec:problem_setting}

Let $\mathcal{D}$ be a document collection, $\mathcal{Q}$ a set of queries,
and $\mathcal{R} \subseteq \mathcal{Q} \times \mathcal{D}$ a set of relevance
judgments. An entity linker $\Lambda$ maps each document $d \in \mathcal{D}$
to a set of entities $\Lambda(d) \subseteq \mathcal{E}$, where $\mathcal{E}$
is a knowledge-base entity vocabulary. Entities are represented by
knowledge-base identifiers. In our experiments, these identifiers correspond
to Wikipedia entities, as required by the WAT~\cite{piccinno2014wat} linker. Mentions that cannot
be linked to a Wikipedia entity are therefore outside the observed entity
set. Unless otherwise stated, unjudged candidate documents contribute to
$df_{\text{cand}}$ but are excluded from $df_{\text{rel}}$ and
$df_{\text{nonrel}}$. Discrimination scores are therefore computed
over the judged portion of the candidate set only.

\smallskip

\noindent\textbf{Task.} Given a query $q$ and a candidate set
$\mathcal{D}_{\text{cand}}(q)$ obtained from a first-stage retrieval method,
re-rank the candidates to produce an ordering where relevant documents are
placed higher, using entity signals derived from $q$ and the linked entities
observed in $\mathcal{D}_{\text{cand}}(q)$.

\smallskip

Although entity-aware rankers differ in how they represent and exploit
entities, they share a common step: selecting which entity signals influence
document ranking. We formalize this as a ranked list of candidate entities
$\hat{E}_q = [e_1, e_2, \ldots]$, where each $e_i \in \mathcal{E}$. We
refer to the component that produces $\hat{E}_q$ as an \emph{entity
selector}. This abstraction unifies several model classes. In
EDRM~\cite{liu-etal-2018-entity}, $\hat{E}_q$ is simply the set of linked
entities, with no separate ranking stage or explicit top-$k$ truncation. In
attention-based models such as Word-Entity
Duet~\cite{xiong2017word}, linked entities are retained but weighted, which
can be viewed as soft selection over $\hat{E}_q$. Retrieval-based approaches
such as EsdRank~\cite{xiong2015esdrank} construct $\hat{E}_q$ by retrieving
related objects from external resources. More recent methods make this step
explicit by learning an entity scoring function and selecting the top-$k$ for
downstream use~\cite{chatterjee2024dreq,chatterjee2025qder}.

For query $q$, let $\mathcal{R}_q$ and $\overline{\mathcal{R}}_q$ denote the
judged relevant and judged non-relevant documents in
$\mathcal{D}_{\text{cand}}(q)$. Let $df_{\text{rel}}(e,q)$ and
$df_{\text{nonrel}}(e,q)$ denote the numbers of judged relevant and judged
non-relevant candidate documents containing entity $e$, and let
$df_{\text{cand}}(e,q)$ denote its total frequency across all candidate
documents, including unjudged ones. In general,
\[
  df_{\text{cand}}(e,q) \geq df_{\text{rel}}(e,q) + df_{\text{nonrel}}(e,q),
\]
with equality only when all candidate documents are judged. The central
question is therefore not simply whether an entity is \emph{about} the query,
but whether it provides useful evidence for ranking documents in this
retrieval setting.
\subsection{Conceptual Entity Relevance (CER)}
\label{sec:cer}

\emph{Conceptual Entity Relevance} (CER) captures whether entity $e$ is
topically related to query $q$, judged from the query text and the entity
description alone, without access to the collection, the entity linker, or
the candidate set. CER corresponds closely to the notion of relevance used by
human judgments, LLM annotations, and other topical supervision signals.

\begin{definition}[CER Label]
A CER label $y^{\textup{cer}}(q,e) \in \{0,1,2\}$ is assigned from the
query title, description, and narrative together with the entity's Wikipedia
title and first-paragraph description: $2$ = core topical match; $1$ =
semantically related but peripheral; $0$ = not relevant or too generic.
\end{definition}

\subsection{Observable Entity Relevance (OER)}
\label{sec:oer}

\emph{Observable Entity Relevance} (OER) asks instead: is observing entity
$e$ in a candidate document associated with higher relevance odds for query
$q$? Unlike CER, which is a property of the query--entity pair alone, OER is
a property of the \emph{query--entity--collection} triple. It depends on how
the entity is linked and distributed across the candidate set, not on topical
relatedness alone.

\begin{definition}[OER Score]
\label{def:oer}
For entity $e$, query $q$, and candidate pool $\mathcal{D}_{\textup{cand}}(q)$,
let
\[
  w(e,q) = 1 - \exp\!\left(-df_{\textup{cand}}(e,q)/\tau\right)
\]
be a soft support weight that downweights entities observed in very few
candidate documents. We define:
\begin{equation}
  \textup{OER}(e,q) =
  w(e,q)\Big[
    \operatorname{logit}\hat{p}(e\mid\textup{rel})
    -
    \operatorname{logit}\hat{p}(e\mid\textup{nonrel})
  \Big],
  \label{eq:oer}
\end{equation}
where
\begin{equation}
  \hat{p}(e\mid\textup{rel}) =
    \frac{df_{\textup{rel}}(e,q)+\alpha}
         {|\mathcal{R}_q|+2\alpha},
  \quad
  \hat{p}(e\mid\textup{nonrel}) =
    \frac{df_{\textup{nonrel}}(e,q)+\alpha}
         {|\overline{\mathcal{R}}_q|+2\alpha}.
  \label{eq:probs}
\end{equation}
We use $\alpha=0.5$ for Jeffreys/additive smoothing and $\tau=5$ as the
support scale parameter. These quantities are smoothed empirical rates, not
calibrated posterior probabilities.
\end{definition}

A positive OER score means that observing $e$ is associated with higher
relevance odds in the candidate set. A score near zero indicates a
non-discriminative entity. A negative score indicates an anti-signal. OER
is computed from relevance judgments and entity linking only, requiring no
entity annotations. It directly reflects the retrieval objective: whether
observed entity evidence discriminates relevant from non-relevant documents.

\subsection{Empirical Hypotheses}
\label{sec:predictions}

The CER--OER distinction leads to three empirical hypotheses that we test in
Sections~\ref{sec:rq1}--\ref{sec:mechanism}.

\smallskip

\noindent\textbf{H1.} CER labels should align only weakly with OER. If the
two notions capture genuinely different properties, agreement should be low
and the disagreement should follow structured failure modes rather than
appearing as random noise.

\smallskip

\noindent\textbf{H2.} CER-based selectors should produce weaker entity
signals than OER-aligned selectors. If CER is insufficient as a supervision
target, CER-trained selectors should prune fewer non-relevant documents,
exhibit poorer score calibration, and face an inescapable
coverage--discrimination tradeoff that cannot be resolved by changing the
entity selection method.

\smallskip

\noindent\textbf{H3.} This mismatch should manifest as a divergence between
closed- and open-world evaluation: CER-based signals may appear effective
when entity evidence is already available, but degrade when the same signals
must operate over the full candidate set.

\section{Experimental Setup}
\label{sec:setup}

\subsection{Data and Retrieval Setup}
\label{sec:dataset}

We experiment on four collections:
(1) \textbf{TREC Robust 2004}~\cite{voorhees2005robust} (249 queries,
${\approx}528$K newswire documents); (2)
\textbf{CODEC}~\cite{mackie2022codec} (42 queries, ${\approx}730$K web
documents) with human-annotated entity relevance labels (CER) for queries;
(3) \textbf{TREC Core 2018}~\cite{allan2017trec,allan2018trec} (50 queries,
${\approx}728$K Washington Post articles); and (4) \textbf{TREC Deep
Learning 2019}~\cite{craswell2020overviewtrec2019deep} (43 queries,
MS~MARCO v1, ${\approx}3.2$M documents).

All collections use Lucene BM25 (default parameters) with RM3 query
expansion as the first-stage retriever ($N=1000$ candidates per query).
RM3 provides a stronger candidate pool than plain BM25, making relevant
documents more likely to appear in the pool.
 Entity linking uses
WAT~\cite{piccinno2014wat}. OER scores and entity ranking analyses are
computed over this same candidate pool.

\subsection{Entity Selectors}
\label{sec:selectors}

We evaluate five test-time-safe entity selectors and two diagnostic upper
bounds while holding the downstream document ranker fixed as QDER~\cite{chatterjee2025qder}. All selectors produce a
top-20 entity ranking per query ($k=20$).

\textbf{PPR Baseline.}
An unsupervised Personalized PageRank~\cite{jeh2003scaling} model
over a local entity graph whose nodes are the union of query-linked
entities and candidate entities, with edges weighted by cosine similarity
between Wikipedia2Vec~\cite{yamada-etal-2020-wikipedia2vec} embeddings. Query-linked entities (from WAT
annotations) serve as seed nodes, weighted by linker confidence. We use this as a lower-bound reference with no learned supervision.

\textbf{BERT (CER, Gemma3).}
A MonoBERT entity ranker trained on LLM-generated CER labels, using
Wikipedia entity descriptions as documents. Labels are generated by
Gemma~3~27B (we also run GPT-4o Mini and find consistent results; we report
Gemma results throughout and prefer the open-source variant). Query--entity
pairs for judgement are drawn from the top-100 PPR entities per query. Each
LLM judges relevance from query text and entity description alone, without
access to corpus statistics or candidate documents.

\textbf{Consensus.}
An unsupervised selector using only the query and the BM25 candidate
set. Its design is motivated by the OER definition: OER rewards entities
that concentrate in relevant documents. Without relevance labels this
cannot be computed directly, but entities that spread uniformly across
the full candidate set are unlikely to be discriminative regardless of
relevance. We therefore select entities by combining support (how
strongly an entity is evidenced across the candidate set) and
specificity (how concentrated it is, via pseudo-IDF over the candidate
pool), requiring each entity to appear in at least two candidate
documents to reduce idiosyncratic linker noise. Concretely,
\begin{equation}
\text{score}(e, q) = \text{soft\_support}(e, q) \times
\left( \log\frac{K+1}{df_{\text{cand}}(e,q)+1} + 1 \right),
\end{equation}
where $K = |\mathcal{D}_{\text{cand}}(q)|$ is the candidate pool size
and the second term is a pseudo-IDF over the candidate pool. We
evaluate three variants differing only in how support is computed:
linker-confidence-weighted (\textit{rho}), BM25-rank-weighted
(\textit{rank}), and combined (\textit{rho+rank}). We report the
best-performing variant (\textit{rho+rank}) throughout due to space
constraints.

\textbf{OER-proxy Listwise.}
A LightGBM LambdaRank model trained on OER log-odds using only
test-time-safe features: candidate-set support statistics
($df_{\text{cand}}$, rank-weighted mention counts, local IDF), stage-1
priors (PPR score, BM25 entity frequency), and lexical overlap. Trained
with 5-fold cross-validation.

\textbf{OER-proxy Pointwise.}
A pointwise LightGBM regressor trained on OER log-odds using the same
features as OER-proxy Listwise. This ablates the training objective within
the OER-proxy family; also evaluated with 5-fold cross-validation.

\textbf{Stats OER Oracle.}
A non-deployable upper bound that ranks entities directly by OER log-odds
computed from document qrels. Evaluated under the conditional protocol only.

\textbf{BERT (CER, DocQrels).}
A MonoBERT entity ranker trained on labels derived from document
relevance judgments. For each query, an entity is labeled positive if
it appears only in relevant documents, negative if it appears only in
non-relevant documents, and excluded if it appears in both. For
closed-world evaluation, the entity ranker is applied to the
qrel-derived entity pool (exclusive positives and negatives), so the
candidate set is restricted to entities with known relevance labels;
the closed-world result therefore reflects oracle-like entity pool
construction and should be interpreted as a supervised upper bound.
For open-world evaluation, the same trained ranker is applied to the
full unfiltered BM25 entity pool without any qrel access, making it a
legitimate test-time evaluation. Architecture and training follow BERT
(CER, Gemma3), differing only in the supervision source.

All supervised models use 5-fold cross-validation over queries,
ensuring that OER scores and entity labels used for training are always
derived from held-out queries. For BERT (CER, Gemma3) and the
OER-proxy models, no ground-truth information from the test fold enters
training or entity selection. For BERT (CER, DocQrels), the open-world
evaluation is test-time-safe; the closed-world evaluation uses a
qrel-derived entity pool and should be interpreted accordingly.
\subsection{OER Operationalizations}
\label{sec:oer_ops}

OER (Eq.~\ref{eq:oer}) requires $df_{\text{rel}}$ and $df_{\text{nonrel}}$
derived from relevance judgments. We operationalize it in two ways.

\textbf{Stats-based OER.} Computed directly from document qrels and WAT
linking output. This serves as both the gold-standard agreement target and
the training signal for the OER-proxy models.

\textbf{LLM-based OER.} Gemma3-27B and GPT-4o Mini are prompted with the
query narrative, candidate-set statistics ($df_{\text{cand}}$,
$df_{\text{rel}}$, $df_{\text{nonrel}}$), and representative candidate-pool
snippets, and asked to assign a label in $\{0,1,2\}$ based on whether the
entity discriminates relevant from non-relevant documents in the provided
evidence. The contrast with BERT (CER, LLM) is exact: the same models and
label scale are used, but the OER version includes corpus evidence while the
CER version does not. We annotate all top-100 entity--query pairs from the
PPR run on each collection. 

\subsection{OER Signal Taxonomy}
\label{sec:taxonomy}

We assign each entity--query pair $(e, q)$ to one of six signal modes:
(1) core signal ($df_{\text{cand}} > 2$, $df_{\text{rel}} \geq 2$,
OER $\geq 0.5$), (2) conditional signal (positive OER below the
core-signal threshold), (3) generic bait (OER $\leq 0$ and
$df_{\text{cand}} \geq 50$), (4) anti-signal (negative OER with
$df_{\text{nonrel}} > df_{\text{rel}}$), (5) sparse evidence
($df_{\text{cand}} \leq 2$), and (6) incidental mention (negative
or zero OER not meeting the generic bait or anti-signal thresholds).

\subsection{Evaluation Metrics}
\label{sec:metrics}

From the OER-based signal labels, we derive several aggregate metrics over
the selected entities. \textbf{Bait rate} is the fraction of top-$k$
selected entities that are generic bait or anti-signal. \textbf{Signal rate}
is the fraction that are core or conditional signals. \textbf{Top-1 bait
rate} is the fraction of queries whose highest-ranked entity is bait.

We also report \textbf{DiscRatio}, a coverage-based metric defined as
$\text{RelCov}/(\text{NonRelCov}+\epsilon)$, where \textbf{RelCov} is the
fraction of relevant candidate documents containing at least one selected
entity, and \textbf{NonRelCov} is defined analogously over non-relevant
documents.

For the Relevance Isolation analysis (Section~\ref{sec:relevance_isolation}),
we additionally report document-filtering metrics over
$\mathcal{D}_{\text{cand}}^{\text{filt}}(q) = \{d \in
\mathcal{D}_{\text{cand}}(q) \mid \Lambda(d) \cap \hat{E}_q^k \neq
\emptyset\}$: \textbf{NonRelPrune}, the fraction of non-relevant candidate
documents removed by the filter, and \textbf{RelRetain}, the fraction of
relevant documents retained.
Although bait rate, signal rate, and DiscRatio are defined relative to OER
(derived from ground truth), this is analogous to evaluating a ranker
against its training target: the metrics measure alignment on held-out
queries.

\subsection{Evaluation Framework}
\label{sec:eval}

The evaluation is staged to trace the CER--OER distinction from label space
to retrieval behavior, corresponding directly to H1--H3.

\textbf{Label Agreement.}
We measure Cohen's $\kappa$ and Pearson/Spearman correlations between CER
labels and multiple OER operationalizations. We also report agreement among
OER operationalizations and a failure-mode breakdown (signal mode $\times$
CER label). Where available, we test whether the pattern holds for native
human CER annotations. This stage evaluates H1.

\textbf{Selection Quality.}
We report bait rate, signal rate, and top-1 bait rate as measures of the
OER quality of selected entities. We also analyze whether a selector's
ranking scores are correlated with discriminative utility: for each selector
we partition entities into ten deciles by rank score and compute mean OER
log-odds per decile. A selector whose scores track discriminative utility
should show a monotonically increasing log-odds curve; a selector whose
scores are uncorrelated with OER should produce a flat curve. This stage
evaluates H2.

\textbf{Relevance Isolation.}
\label{sec:relevance_isolation}
We measure how effectively selected entities isolate relevant documents
from the candidate set via NonRelPrune and RelRetain. A selector
that retains $97\%$ of relevant documents while pruning only $11\%$ of
non-relevant ones offers little practical benefit regardless of entity-level
accuracy. This stage further evaluates H2.

\textbf{Closed- and Open-World Evaluation.}
Closed-world evaluation ranks over $\mathcal{D}_{\text{cand}}^{\text{filt}}(q)$
and measures how effectively a model exploits entity evidence once available.
Open-world evaluation ranks over the full candidate pool
$\mathcal{D}_{\text{cand}}(q)$ and assesses whether those signals remain
useful without reachability assumptions. We report both as they can diverge:
CER-based selectors may perform well when entity evidence is present but
transfer poorly to full candidate-set ranking. Open-world MAP is the primary
metric for test-time-safe selectors; closed-world MAP is reported for
diagnosis. Selectors requiring relevance labels at selection time are
evaluated closed-world only. This stage evaluates H3.

\section{The CER--OER Gap}
\label{sec:rq1}

We first test H1: if CER and OER are genuinely different notions of
entity usefulness, their agreement should be weak and the disagreement
should be structured. We focus on the Label Agreement stage of the
evaluation framework (Section~\ref{sec:eval}), beginning with Robust04
and then testing whether the same pattern holds across additional
collections. Unless otherwise stated, detailed failure-mode breakdowns
use Gemma3-27B as the default LLM-based OER judge; other OER variants
show consistent trends.

\subsection{Label Agreement on Robust04}
\label{sec:rq1_robust04}

\begin{table}[t]
\centering
\small
\caption{\textbf{CER--OER agreement across collections.}
All values are significant at $p<0.001$. Samples: Robust04 ($n=24{,}868$), DL19 ($n=4{,}300$),
CORE18 ($n=5{,}000$), CODEC ($n=4{,}200$).
GPT-4o Mini labels only available for Robust04.
The final block reports an additional CODEC-only comparison using native
human entity qrels as the CER source
($n=5{,}751$); both Stats OER and Gemma3 OER are included as OER
targets to show the pattern holds across OER operationalizations.}
\label{tab:agreement_all}
\scalebox{0.78}{
\begin{tabular}{lcccc}
\toprule
\textbf{Measure} & \textbf{Robust04} & \textbf{DL19} & \textbf{CORE18} & \textbf{CODEC} \\
\midrule
$\kappa(\text{CER},\, \text{Stats OER})$            & 0.062 & 0.093 & 0.085 & 0.091 \\
$\kappa(\text{CER},\, \text{Gemma3 OER})$            & 0.171 & 0.232 & 0.175 & 0.059 \\
$\kappa(\text{CER},\, \text{GPT-4o Mini OER})$       & 0.236 & --    & --    & --    \\
$\kappa(\text{Stats OER},\, \text{Gemma3 OER})$      & 0.483 & 0.476 & 0.524 & 0.574 \\
$\kappa(\text{Gemma3 OER},\, \text{GPT OER})$        & 0.441 & --    & --    & --    \\
\midrule
Pearson $r(\text{CER},\, \text{OER log-odds})$       & 0.209 & 0.136 & 0.091 & 0.110 \\
Spearman $\rho(\text{CER},\, \text{OER log-odds})$   & 0.174 & 0.115 & 0.080 & 0.084 \\
\midrule
\multicolumn{5}{l}{\textit{Additional CODEC check: native human entity qrels as CER source ($n=5{,}751$)}} \\[2pt]
$\kappa(\text{Human CER},\, \text{Stats OER})$                    & -- & -- & -- & 0.076 \\
$\kappa(\text{Human CER},\, \text{Gemma3 OER})$                   & -- & -- & -- & 0.043 \\
Pearson $r(\text{Human CER},\, \text{OER log-odds})$              & -- & -- & -- & 0.157 \\
Spearman $\rho(\text{Human CER},\, \text{OER log-odds})$          & -- & -- & -- & 0.108 \\
\bottomrule
\end{tabular}
}
\end{table}

Table~\ref{tab:agreement_all} shows that CER and OER agree only weakly.
On Robust04, agreement between CER and stats-based OER is near chance
($\kappa=0.062$), and remains low for LLM-based OER ($\kappa=0.171$
with Gemma3 and $0.236$ with GPT-4o Mini). By contrast, OER
operationalizations agree substantially better with one another
($\kappa=0.483$ between stats-based and Gemma3 OER; $\kappa=0.441$
between Gemma3 and GPT-based OER). CER is the outlier among these relevance signals.
Correlation analysis confirms this: CER exhibits only a weak
relationship with OER log-odds ($r=0.209$, $\rho=0.174$), indicating
that semantic relatedness is a poor proxy for observable discriminative
utility.

\subsection{The Disagreement Is Structured}
\label{sec:rq1_structure}

Weak overall agreement does not explain why CER fails as a supervision
target. The key question is whether the disagreement reflects random
noise or a systematic mismatch between topical plausibility and
retrieval utility. To examine this, we cross-tabulate CER labels against
OER signal modes for all query--entity pairs. We report results on
Robust04 ($24{,}868$ pairs); other datasets show the same pattern.
Three trends stand out.

\smallskip
\noindent\textbf{CER overconfidence produces weak and non-signals.}
Among the 992 pairs assigned the highest CER label
($y^{\textup{cer}}=2$), only $24.1\%$ are classified as strong
OER signals. A further $50.9\%$ are judged only weakly positive,
and $25.0\%$ are judged non-signals.

\smallskip
\noindent\textbf{CER misses discriminative entities.}
Among the $20{,}991$ pairs with $y^{\text{cer}}=0$, Gemma3-27B assigns
positive OER labels to 27.5\%, including $1.6\%$
classified as strong OER signals. CER labels these as non-relevant
despite observable evidence of retrieval utility.

\smallskip
\noindent\textbf{Disagreement concentrates in high-frequency entities.}
Table~\ref{tab:dfcand_breakdown} breaks down the CER--OER disagreement
by entity candidate-set frequency ($df_{\text{cand}}$), pooled across
all four collections. Among entities with $df_{\text{cand}}>50$,
$22.2\%$ of CER label~2 entities are OER non-signals---the highest
overconfidence rate across all frequency bins. These are precisely the
high-frequency entities such as \textsc{United States} that a CER-based
judge correctly identifies as topically related but that the linker
fires on indiscriminately across relevant and non-relevant documents.
The CER=0$\to$signal rate remains high across all bins
($41\%$--$70\%$), confirming that CER misses discriminative entities at
every frequency level, not only among rare entities.

\begin{table}[t]
\centering
\small
\caption{CER--OER disagreement by entity candidate-set frequency
($df_\text{cand}$), pooled across all four collections. Pairs with
$df_{\text{cand}}=0$ are excluded from the frequency breakdown
($n=36{,}020$ of $38{,}368$ total pairs).
CER=2$\!\to\!$non-signal: fraction of top-confidence CER entities
classified as OER non-signals (generic bait, anti-signal, or
incidental mention). CER=0$\!\to\!$signal: fraction of CER-rejected
entities classified as OER-positive signals.}
\label{tab:dfcand_breakdown}
\begin{tabular}{lrrr}
\toprule
$df_\text{cand}$ & $N$ & CER=2$\!\to\!$non-signal\,\% &
CER=0$\!\to\!$signal\,\% \\
\midrule
1--5    &  1,082 & 19.0 & 40.7 \\
6--50   &  7,496 & 21.3 & 69.7 \\
${>}50$ & 27,442 & 22.2 & 55.3 \\
\bottomrule
\end{tabular}
\end{table}

\smallskip
\noindent\textbf{Qualitative example from Robust04.}
For the query \emph{Thatcher resignation impact} (query~666),
\textsc{Margaret Thatcher} receives CER label~2 yet its stats-based OER
log-odds is $-3.03$ ($df_{\text{rel}}=2$, $df_{\text{nonrel}}=360$).
Non-relevant documents mention her only as a reference point, while
relevant documents analyze the political consequences of her departure.
The entity is topically central but not discriminative in this candidate
pool. Conversely, for the query \emph{Russia--Cuba economic relations}
(query~617), \textsc{Soviet Union} receives CER label~0 yet has OER
log-odds $=+4.53$ ($df_{\text{rel}}=62$, $df_{\text{nonrel}}=94$).
Relevant documents discuss dissolved Soviet economic and military ties
with Cuba directly, while non-relevant documents mention the Soviet
Union only in passing. A human CER judgment labels this entity as not
relevant, but its observed distribution makes it one of the strongest
signals in the candidate pool. Additional examples are provided in the Github repository.

\smallskip
\noindent\textbf{Qualitative example from CODEC.}
CODEC provides human entity relevance judgements, allowing us to
validate the CER--OER mismatch under human rather than LLM-generated
CER labels; OER labels are stats-based throughout. For the query
\emph{Why are some economists sceptical about the EU's monetary union
without a shared fiscal system?} (\texttt{economics-6}), human
annotators assign \textsc{Greek withdrawal from the eurozone} CER
label~0 yet its OER log-odds is $+2.20$ ($df_{\text{rel}}=3$,
$df_{\text{nonrel}}=0$): every document containing this entity discusses
the prospect of Greek exit from the eurozone directly, making it one of
the strongest discriminative signals available despite being labeled
non-relevant on topical grounds.

\smallskip
Together, these results show that CER favors semantic plausibility over
discriminative utility. It assigns high labels to many entities that are
weak or non-discriminative in the candidate pool. It also labels many
useful discriminative entities as non-relevant. These patterns are
consistent across LLM-based OER variants on Robust04; GPT-4o Mini
yields the same qualitative conclusions ($\kappa=0.236$).

\subsection{Generalization Across Collections}
\label{sec:rq1_generalization}

The same pattern holds across other collections
(Table~\ref{tab:agreement_all}). Agreement between CER and OER remains
consistently low, ranging from $\kappa=0.062$ to $0.093$ with
stats-based OER and from $0.059$ to $0.232$ with Gemma3. Agreement
among OER operationalizations is substantially higher throughout,
reaching $\kappa=0.574$ on CODEC. On CODEC, agreement drops to
near-zero ($\kappa=0.059$) while OER operationalizations exhibit their
strongest mutual agreement---further reinforcing that CER is the
systematic outlier across domains.

The structured failure modes also generalize. Among entities with CER
label~2, the non-signal rate is $24.7\%$ on DL19 and $23.1\%$ on
Core18, close to the $25.0\%$ rate on Robust04 (all figures using
stats-based OER). CER also continues to miss discriminative entities:
$39.6\%$ of CER label~0 entities are OER-positive on DL19 and $47.0\%$
on Core18, compared to $23.4\%$ on Robust04 (stats-based OER). The
corresponding figure on Robust04 using Gemma3-based OER is $27.5\%$,
as reported in Section~\ref{sec:rq1_structure}; the difference reflects
the two OER operationalizations rather than an inconsistency. The
increasing rate across collections suggests the gap widens in denser
or noisier entity-linking environments.

Using CODEC's human entity relevance judgements as CER labels, we
measure agreement with both stats-based and LLM-based OER across all
$5{,}751$ query--entity pairs in the intersection of the human qrel set
and OER statistics. Agreement remains weak: $\kappa=0.076$ against
stats-based OER and $\kappa=0.043$ against Gemma3 OER ($r=0.157$,
$\rho=0.108$). The CER--OER gap is therefore not an artifact of LLM
prompting or labeling style; it persists under human CER judgments and
across both statistical and LLM-based OER operationalizations. A
companion study providing extended evidence across 443 configurations of unsupervised entity-oriented document ranking methods
suggests that this ceiling is a structural property of the linking
environment rather than a model-specific
artifact~\cite{chatterjee2026entitiesretrievalsignalssystematic}.

\subsection*{Answer to H1}

CER and OER exhibit weak agreement across all evaluated settings, while
OER operationalizations agree with one another substantially more
strongly. The disagreement is systematic: CER overvalues semantically
plausible entities and misses entities whose observed links distinguish
relevant from non-relevant documents. This pattern holds across
datasets, annotation sources, and OER operationalizations. The
CER--OER gap is therefore large, structured, and general. We next ask
whether this label-space mismatch affects the entities selected for
ranking.

\section{Consequences of the CER--OER Gap}
\label{sec:rq2}

Section~\ref{sec:rq1} showed that CER and OER are weakly aligned. We now test H2: if CER is an insufficient retrieval supervision target, this mismatch should appear in learned selector behavior. Specifically, CER-trained selectors should favor semantically plausible but weakly discriminative entities, producing low score--discriminativeness correlation and higher bait rates. We test this through the Selection Quality and Relevance Isolation stages of our evaluation framework.

\subsection{Selection Quality}
\label{sec:rq2_levelb}

\begin{table}[t]
  \centering
  \caption{Entity run diagnostics at $k=20$ on Robust04. Bait rate,
signal rate, and top-1 rates measure OER quality of selections.
RelCov and NonRelCov are the fractions of relevant and non-relevant
candidate documents containing at least one top-20 entity
(exact document-level matching, macro-averaged over queries);
DiscRatio $= \text{RelCov} / \text{NonRelCov}$.}
  \label{tab:entity_diagnostics}
  \setlength{\tabcolsep}{4pt}
  \scalebox{0.75}{
  \begin{tabular}{lccccc}
    \toprule
    System
      & Bait$\downarrow$ & Signal$\uparrow$
      & Top1-Bait$\downarrow$ & Top1-Sig$\uparrow$
      & DiscRatio$\uparrow$ \\
    \midrule
    PPR Baseline          & 0.406 & 0.567 & 0.277 & 0.707 & 1.045 \\
    BERT (CER, Gemma3)    & 0.292 & 0.630 & 0.169 & 0.787 & 1.094 \\
    Consensus (rho+rank)  & 0.213 & 0.712 & 0.196 & 0.744 & 1.056 \\
    OER-proxy Listwise    & 0.150 & 0.735 & 0.080 & 0.859 & 1.205 \\
    \midrule
    Stats OER Oracle      & 0.075 & 0.850 & 0.020 & 0.952 & 5.403 \\
    BERT (CER, DocQrels)  & 0.002 & 0.716 & 0.000 & 0.751 & 154.9 \\
    \bottomrule
  \end{tabular}
  }
\end{table}

Table~\ref{tab:entity_diagnostics} reports Selection Quality metrics
across the full supervision ladder on Robust04. BERT (CER, Gemma3)
yields a bait rate of $0.292$: nearly one in three top-20 selections is
generic bait or an anti-signal, and for $16.9\%$ of queries the
top-ranked entity is bait. Its signal rate is only $0.630$. Consensus
already improves on this with a lower bait rate ($0.213$) and higher
signal rate ($0.712$), showing that simple corpus-based heuristics
outperform CER-based supervision for discriminative entity selection.

At the aggregate level, BERT (CER, Gemma3) achieves DiscRatio $= 1.094$,
making its selected entities effectively indistinguishable from random
for relevant--non-relevant discrimination. Its non-zero signal rate
therefore reflects weak local utility that does not translate into
meaningful aggregate discrimination. The extreme DiscRatio reported for
BERT (CER, DocQrels) ($154.9$) is not a general property of that
selector: it arises because the DocQrels variant retains only entities
occurring exclusively in relevant or non-relevant documents, effectively
acting as an oracle-like filter under conditional analysis. As we show
in Section~\ref{sec:rq3}, this behavior does not transfer to open-world
retrieval.

\begin{figure}[t]
  \centering
  \includegraphics[scale=0.28]{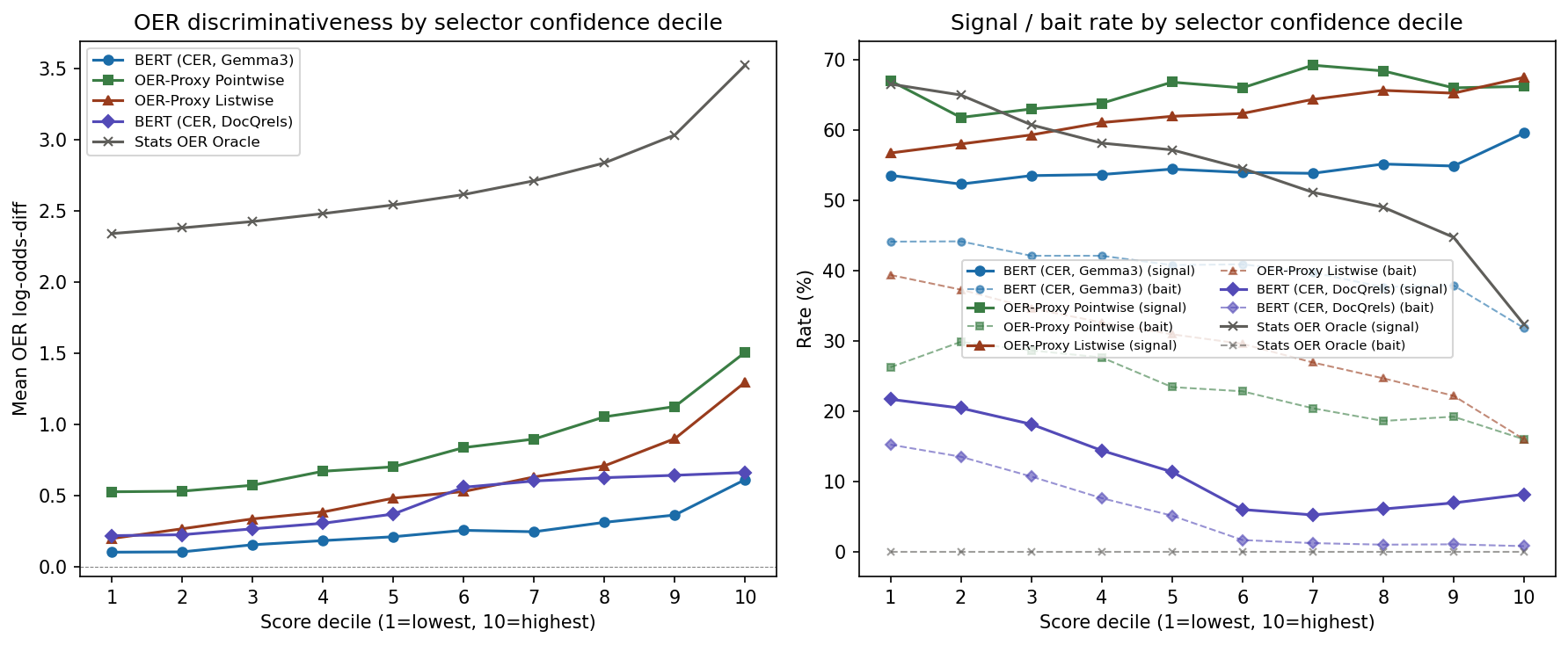}
  \caption{OER discriminativeness (mean log-odds-diff, left) and
  signal/bait rates (right) by selector confidence decile on Robust04.
  BERT (Gemma3) shows a near-flat log-odds curve ($0.10 \to 0.61$).
  OER-proxy selectors are monotonically increasing. BERT (DocQrels) ($0.22 \to 0.66$) barely exceeds
  BERT (Gemma3) despite perfect CER supervision.}
  \label{fig:decile_curves}
\end{figure}

We further examine whether selector scores correlate with discriminative
utility via a score-decile analysis on Robust04. CER-based scores are
largely uncorrelated with corpus discriminativeness: a flat CER curve
means the top-ranked entities are no more informative than the
bottom-ranked ones, as observed for BERT (CER, Gemma3). The limited
improvement of BERT (CER, DocQrels) despite near-perfect CER
supervision confirms that the issue is not label noise, but that CER
score does not predict OER quality. The same pattern holds across
collections: on Core18, BERT (CER, Gemma3) yields bait rate $=0.342$
and signal rate $=0.567$; on DL19, $0.307$ and $0.631$.

\smallskip
\noindent\textbf{Qualitative example.}
For the query \emph{Why is universal healthcare so politically
controversial in the United States?} (CODEC, \texttt{politics-12}),
human annotators assign \textsc{Universal Health Care} (the
query topic itself) CER label~2, yet its OER log-odds is $-0.39$
($df_{\text{rel}}=13$, $df_{\text{nonrel}}=13$ among judged documents,
$df_{\text{cand}}=144$). The entity appears in equal numbers of judged
relevant and judged non-relevant documents, but is broadly distributed
across the full candidate pool. Its negative OER score reflects that
the judged non-relevant pool for this query is smaller than the relevant
pool, making the entity's observed presence rate higher among non-relevant
documents despite equal raw counts. Both the stats-based and LLM-based
OER judges classify it as generic bait. A CER-based selector ranks it
at the top; an OER selector correctly identifies it as non-discriminative.

\subsection{Relevance Isolation}
\label{sec:relevance_isolation}

\begin{table}[t]
  \centering
  \caption{Document-level effects of entity signals at $k=20$
  (5-fold CV means). Higher is better for both metrics.}
  \label{tab:doc_filter}
  \setlength{\tabcolsep}{5pt}
  \scalebox{0.85}{
  \begin{tabular}{llcc}
    \toprule
    Collection & System
      & RelRetain$\uparrow$ & NonRelPrune$\uparrow$ \\
    \midrule
    \multirow{3}{*}{Robust04}
      & BERT (CER, Gemma3)     & 0.971 & 0.113 \\
      & OER-Proxy (Pointwise)  & 0.963 & 0.191 \\
      & OER-Proxy (Listwise)   & 0.956 & 0.212 \\
    \midrule
    \multirow{3}{*}{Core18}
      & BERT (CER, Gemma3)     & 0.994 & 0.022 \\
      & OER-Proxy (Pointwise)  & 0.957 & 0.150 \\
      & OER-Proxy (Listwise)   & 0.889 & 0.233 \\
    \midrule
    \multirow{3}{*}{DL19}
      & BERT (CER, Gemma3)     & 0.991 & 0.038 \\
      & OER-Proxy (Pointwise)  & 0.976 & 0.088 \\
      & OER-Proxy (Listwise)   & 0.973 & 0.098 \\
    \midrule
    \multirow{3}{*}{CODEC}
      & BERT (CER, Gemma3)     & 0.905 & 0.136 \\
      & OER-Proxy (Pointwise)  & 0.933 & 0.086 \\
      & OER-Proxy (Listwise)   & 0.386 & 0.612 \\
    \bottomrule
  \end{tabular}
  }
\end{table}

Table~\ref{tab:doc_filter} reports Relevance Isolation metrics across
all four collections, measuring how useful the selected entities are for
isolating relevant documents from the candidate set.

On Robust04, BERT (CER, Gemma3) preserves $97.1\%$ of relevant documents but removes only $11.3\%$ of non-relevant ones, leaving an average of $850.7$ candidates per query. The effect is weaker on Core18 and DL19, where NonRelPrune drops to $0.022$ and $0.038$, respectively. Thus, the CER-trained entity channel provides little effective signal for filtering non-relevant documents.

OER-proxy Listwise consistently improves non-relevant pruning over BERT
(CER, Gemma3) across three collections: $+0.099$ on Robust04 ($0.212$
vs.\ $0.113$), $+0.211$ on Core18 ($0.233$ vs.\ $0.022$), and $+0.060$
on DL19 ($0.098$ vs.\ $0.038$). The pattern is consistent: the weaker
the CER-based filtering, the larger the gain from OER alignment,
suggesting a systematic consequence of supervision mismatch rather than
a collection-specific artifact.

CODEC is the exception: OER-proxy Listwise achieves high NonRelPrune ($0.612$), but RelRetain falls to $0.386$. We attribute this to instability from CODEC's small query set (42 queries) and sparser entity linking in web documents than in newswire. This exposes a real failure mode of OER alignment: with small candidate pools and sparse linking, OER supervision can overfit to discriminative but rare entities, improving non-relevant pruning at the cost of relevant-document coverage.

\subsection{Coverage--Discrimination Tradeoff}
\label{sec:coverage_ceiling}

The Relevance Isolation results show that CER-based selectors fail to
prune non-relevant documents. A natural question is whether this can be
fixed by selecting different entities or cleaning the signal post-hoc.
Both approaches fail for structural reasons.

We evaluate 193 unsupervised entity selection configurations 
on Robust04,
spanning similarity-based, graph-based, LLM-based, and hybrid methods
with variation in embedding source, similarity function, and aggregation
strategy. RelCov and NonRelCov are nearly perfectly correlated
($r=0.954$): every gain in relevant-document coverage brings a
proportional increase in non-relevant coverage, regardless of method
family. No configuration achieves high RelCov with low NonRelCov. The
coverage--discrimination tradeoff is a structural property of the entity
linking environment, not a tuning problem.


We next test whether post-hoc signal cleaning can escape this tradeoff. Using stats-based OER as an oracle filter, we remove entities below increasing thresholds $\tau$ from the BERT (CER, Gemma3) ranking and pass the remaining entities to a fixed downstream ranker. Table~\ref{tab:oer_filtering} shows that filtering improves entity-level discrimination: DiscRatio rises from $1.219$ to $1.769$, while NonRelCov falls from $0.735$ to $0.373$. However, this gain comes through coverage loss. RelCov drops from $0.895$ to $0.660$, making $34\%$ of relevant documents structurally unreachable at the strictest threshold. Downstream MAP remains poor ($0.210$--$0.232$), below both the unfiltered run ($0.310$) and BM25 ($0.292$). Thus, coverage loss dominates: once broad-coverage entities are removed, the entity channel provides less signal than term-based retrieval. Post-hoc signal cleaning cannot resolve the tradeoff.

\begin{table}[t]
\centering
\small
\caption{Post-hoc OER filtering applied to the BERT (CER, Gemma3)
entity run at increasing thresholds $\tau$, with the document ranker
held fixed. RelCov and NonRelCov measure the fractions of relevant
and non-relevant candidate documents retained after filtering to
entity-matched documents (macro-averaged over queries). Discrimination
improves with $\tau$ but relevant-document coverage drops
monotonically, and downstream ranking falls below both the
unfiltered baseline and BM25+RM3 at every threshold.}
\label{tab:oer_filtering}
\scalebox{0.75}{
\begin{tabular}{lcccccc}
\toprule
\textbf{Run} & \textbf{RelCov} & \textbf{NonRelCov} &
\textbf{DiscRatio} & \textbf{MAP} & \textbf{nDCG@20} & \textbf{P@20} \\
\midrule
BM25+RM3 (baseline)     & ---   & ---   & ---   & 0.292 & 0.435 & 0.383 \\
LLM CER (unfiltered)    & 0.971 & 0.888 & 1.09  & 0.310 & 0.464 & 0.404 \\
\midrule
OER filtered $\tau=0.0$ & 0.895 & 0.735 & 1.219 & 0.210 & 0.352 & 0.307 \\
OER filtered $\tau=0.5$ & 0.797 & 0.560 & 1.423 & 0.221 & 0.373 & 0.328 \\
OER filtered $\tau=1.0$ & 0.660 & 0.373 & 1.769 & 0.232 & 0.386 & 0.337 \\
\bottomrule
\end{tabular}
}
\end{table}

\subsection*{Answer to H2}

CER-trained selectors produce low-quality entity runs: high bait rates,
weak aggregate discrimination, and near-flat calibration curves confirm
that their confidence scores carry little information about corpus
discriminativeness. These differences propagate into document
filtering---CER-based selectors prune very few non-relevant documents,
while OER-aligned selectors consistently improve pruning across
collections. This failure is structural. The coverage--discrimination
tradeoff cannot be escaped by selecting different entities or cleaning
the signal post-hoc (Section~\ref{sec:coverage_ceiling}). The
supervision target is the decisive factor. We next ask whether this
degradation in entity signal quality translates into downstream
retrieval losses.

\section{OER Supervision Improves Retrieval}
\label{sec:rq3}

Sections~\ref{sec:rq1} and~\ref{sec:rq2} showed that CER and OER diverge at the label level, and that this mismatch weakens entity ranking. We now test H3: CER-based signals may help when entity reachability is assumed, but should transfer poorly when the entity channel must operate over the full candidate set. We evaluate this by running QDER with each entity selector and reporting downstream MAP on Robust04. Section~\ref{sec:rq2} showed that the same entity-level trends generalize across collections.

\subsection{The Open-World Ceiling Is General}
\label{sec:rq3_signature}

\begin{table}[t]
  \centering
  \caption{Closed- vs.\ open-world MAP and nDCG@20 on Robust04
  across six entity-oriented architectures.
  All models collapse to near-BM25 under open-world evaluation
  regardless of closed-world performance.}
  \label{tab:arch_comparison}
  \setlength{\tabcolsep}{4pt}
  \scalebox{0.8}{
  \begin{tabular}{lcccc}
    \toprule
    & \multicolumn{2}{c}{Closed-world}
    & \multicolumn{2}{c}{Open-World} \\
    \cmidrule(lr){2-3}\cmidrule(lr){4-5}
    Model & MAP & nDCG@20 & MAP & nDCG@20 \\
    \midrule
    BM25+RM3         & ---   & ---   & 0.292 & 0.435 \\
    \midrule
    EDRM-KNRM        & 0.089 & 0.153 & 0.089 & 0.151 \\
    EDRM-ConvKNRM    & 0.087 & 0.150 & 0.088 & 0.150 \\
    Word-Entity Duet & 0.151 & 0.241 & 0.149 & 0.235 \\
    EVA              & 0.156 & 0.312 & 0.167 & 0.302 \\
    EsdRank          & 0.124 & 0.213 & 0.111 & 0.202 \\
    DREQ             & 0.697 & 0.867 & 0.293 & 0.439 \\
    QDER             & 0.608 & 0.769 & 0.294 & 0.438 \\
    \bottomrule
  \end{tabular}
  }
\end{table}

Before isolating the effect of supervision, we establish that the
open-world performance ceiling is a general phenomenon rather than an
artifact of a particular architecture. Table~\ref{tab:arch_comparison}
compares closed- and open-world MAP across six entity-oriented neural
architectures on Robust04. DREQ~\cite{chatterjee2024dreq} achieves
closed-world MAP $=0.697$ but drops to $0.293$ open-world;
QDER~\cite{chatterjee2025qder} drops from $0.608$ to $0.294$. Both
collapse to the BM25+RM3 baseline. Earlier
architectures---EDRM~\cite{liu-etal-2018-entity}, Word-Entity
Duet~\cite{xiong2017word}, EVA~\cite{tran2022dense}, and
EsdRank~\cite{xiong2015esdrank}---show little or no closed-/open-world
gap and neither setting improves over BM25. Across all six
architectures, no model improves open-world MAP by more than $0.002$
over BM25+RM3.

To confirm that the failure originates on the entity side rather than
from the expansion of the document pool, we vary the entity side and
document side independently using DREQ~\cite{chatterjee2024dreq}---chosen
over QDER to confirm the finding is not architecture-specific. Holding
the entity side closed-world, MAP stays at $0.695$--$0.698$ regardless
of document-side filtering. Moving the entity side to open-world, MAP
collapses to $0.291$--$0.293$ regardless of document-side filtering.
The document-side decision contributes at most $0.003$ MAP points; the
entity-side decision contributes $0.405$. The failure is therefore localized to the entity side, rather than to the
expansion of the document pool.

This ceiling is not limited to the six supervised architectures. A companion study evaluates 443 unsupervised entity-oriented ranking configurations on Robust04, covering embedding-based, graph-based, LLM-based, and hybrid selection methods with varied weighting and scoring choices. Under open-world evaluation, none surpasses BM25+RM3: mean MAP is $0.231$, median MAP is $0.241$, and the maximum is $0.289$~\cite{chatterjee2026entitiesretrievalsignalssystematic}. Thus, the open-world ceiling persists across both architectures and supervision strategies.

\subsection{Fixing the Ranker Isolates Supervision}
\label{sec:rq3_fixed_ranker}

\begin{table}[t]
  \centering
  \caption{Downstream QDER MAP and nDCG@20 on Robust04 under
closed- and open-world evaluation. The document ranker is held
fixed across all rows; only the entity selector changes.
$\dagger$ denotes statistically significant improvement over
BM25+RM3 ($p < 0.05$, paired $t$-test). Stats OER Oracle
requires qrel access at selection time and has no open-world
variant (---). BERT (CER, DocQrels) uses a qrel-derived entity
pool for closed-world evaluation only; its open-world result
uses the full unfiltered BM25 entity pool and is a legitimate
test-time evaluation.}
  \label{tab:downstream}
  \setlength{\tabcolsep}{4pt}
  \scalebox{0.7}{
  \begin{tabular}{llcccc}
    \toprule
    Entity Selector & Selector type
      & \multicolumn{2}{c}{Closed-world}
      & \multicolumn{2}{c}{Open-world} \\
    \cmidrule(lr){3-4}\cmidrule(lr){5-6}
    & & MAP & nDCG@20 & MAP & nDCG@20 \\
    \midrule
    BM25+RM3          & No entity       & ---   & ---   & 0.292 & 0.435 \\
    GEEER  & CER (unsup)     & ---   & ---   & 0.275 & 0.408 \\
    BERT (CER, DocQrels) & CER (oracle)    & 0.608$\dagger$ & 0.769$\dagger$ & 0.294 & 0.438 \\
    BERT (CER, Gemma3)    & CER (LLM)       & 0.310$\dagger$ & 0.438 & 0.310$\dagger$ & 0.464$\dagger$ \\
    Consensus         & CER (unsup)     & 0.306 & 0.457$\dagger$ & 0.306 & 0.455$\dagger$ \\
    \midrule
    OER-proxy Listwise    & OER (learned)   & 0.333$\dagger$ & 0.461$\dagger$ & 0.343$\dagger$ & 0.491$\dagger$ \\
    \midrule
    Stats OER Oracle  & OER (oracle)    & 0.396$\dagger$ & 0.570$\dagger$ & ---   & --- \\
    \bottomrule
  \end{tabular}
  }
\end{table}

To obtain controlled evidence for the effect of supervision, we hold the
downstream document ranker fixed as QDER and vary only the entity
selector. Because the ranker, candidate pool, and architecture are
identical across all rows of Table~\ref{tab:downstream}, any differences
in retrieval performance are attributable to the entity selector. Using
OER-proxy Listwise as the selector, QDER achieves open-world MAP
$=0.343$---the highest among all test-time-safe selectors, a gain of
$+0.051$ over BM25+RM3 and $+0.033$ over BERT (CER, Gemma3). It
remains competitive with or above several strong neural rerankers
including RankT5~\cite{zhuang2023rankt5} ($0.303$), MonoBERT~\cite{nogueira2019passage} ($0.297$), and RankZephyr~\cite{pradeep2023rankzephyr}
($0.318$). Two findings stand out.

\smallskip
\noindent\textbf{OER improves over CER supervision end-to-end.}
Using OER-proxy Listwise, QDER reaches MAP $=0.343$ under open-world
evaluation; using BERT (CER, Gemma3) it reaches $0.310$. Both use the
same candidate pool and downstream architecture. The difference is
solely the supervision target used to produce the entity run, isolating
the effect of supervision on downstream ranking.

\smallskip
\noindent\textbf{Unsupervised CER-style methods do not close the gap.}
To test whether the limitation stems from imperfect supervision rather
than the CER objective itself, we include
GEEER~\cite{gerritse2020geeer}, a strong unsupervised entity ranker that uses pre-trained
Wikipedia2Vec~\cite{yamada-etal-2020-wikipedia2vec} embeddings without
relying on relevance labels. GEEER ranks entities by semantic similarity
to the query---a clean CER-style selection baseline---with no label
noise. Using GEEER as the selector, QDER achieves open-world MAP
$=0.275$, below BM25+RM3 ($0.292$). Using Consensus, QDER reaches
$0.306$, slightly above BM25+RM3 but still below BERT (CER, Gemma3).
Removing supervision does not resolve the gap---the limitation lies in
the CER-style selection objective itself.
Under open-world evaluation, CER-based selectors with corpus or
label supervision (Consensus and BERT (CER, Gemma3)) cluster between
$0.306$ and $0.310$. The purely semantic selector GEEER reaches only
$0.275$, below BM25. The improvement from the best supervised
CER-based selector to OER-proxy Listwise ($+0.033$) substantially
exceeds the variation between Consensus and BERT (CER, Gemma3) ($0.004$),
indicating a qualitative difference between OER and CER supervision
regimes.

\subsection*{Answer to H3}

OER-aligned supervision improves retrieval precisely where predicted:
under open-world evaluation, where the entity channel must operate over
the full candidate set. CER-based selectors, across architectures and
supervision sources, cluster near the BM25+RM3 baseline. The gap between OER and supervised CER regimes substantially exceeds
the variation within supervised CER selectors, indicating
a qualitative rather than a marginal difference. The bottleneck lies not
in the ranker architecture but in the quality of the entity signals
provided to it.

\section{Why OER Alignment Improves Retrieval}
\label{sec:mechanism}

Section~\ref{sec:rq3} showed that OER-aligned supervision improves
open-world retrieval when the document ranker is held fixed. We now ask
why. The key point is not that OER-proxy introduces novel features, but
that it aligns entity selection with the retrieval objective. We
disentangle two factors: the candidate-set features available to the
selector, and the supervision target used to train it.

\begin{table}[t]
\centering
\small
\caption{Feature ablation and supervision swap results on Robust04 (5-fold CV means). \emph{Semantic only}: OER supervision with only semantic features. \emph{+Candidate support}: adds $df_\text{cand}$, rank-weighted counts, local IDF. \emph{+Stage-1 priors}: adds PPR and BM25 priors. \emph{CER supervision (swap)}: full feature set with CER as training target. Reference rows in italics.}
\label{tab:ablation_docfilter}
\scalebox{0.8}{
\begin{tabular}{lrr}
\toprule
System & RelRetain\,$\uparrow$ & NonRelPrune\,$\uparrow$ \\
\midrule
\textit{BERT (CER) [ref]} & \textit{0.971} & \textit{0.113} \\
\midrule
Semantic only & 0.975 & 0.092 \\
+Candidate support & 0.934 & 0.261 \\
+Stage-1 priors & 0.957 & 0.215 \\
CER supervision (swap) & 0.964 & 0.130 \\
\midrule
\textit{OER-proxy LTR (full) [ref]} & \textit{0.956} & \textit{0.212} \\
\bottomrule
\end{tabular}
}
\end{table}

\smallskip
\noindent\textbf{Candidate-set evidence provides the main signal.}
Table~\ref{tab:ablation_docfilter} reports ablations with the
supervision target fixed to OER log-odds. Semantic features alone (BERT
entity score and lexical overlap) achieve NonRelPrune $=0.092$, below
BERT (CER, Gemma3) ($0.113$) despite using OER supervision. These
features capture query--entity relatedness but not how entities are
distributed across relevant and non-relevant documents. Adding
candidate-set support statistics ($df_{\text{cand}}$, rank-weighted
mention counts, local IDF) raises NonRelPrune to $0.261$---a $183\%$
relative improvement over semantic-only and $131\%$ over BERT (CER,
Gemma3)---while RelRetain remains high ($0.934$). Adding stage-1 priors
or the full feature set does not improve this pruning score ($0.215$ and
$0.212$, respectively), although the full model is used for downstream
ranking. Candidate-set support is therefore the main source of the gain.

\smallskip
\noindent\textbf{OER supervision alone is not enough.}
Training on OER log-odds using only semantic features yields NonRelPrune
$=0.092$, below BERT (CER, Gemma3) ($0.113$). The correct supervision
target cannot compensate for features that lack discriminative evidence.
This shows that OER alignment requires observable candidate-set signals,
not only a different training label.

\smallskip
\noindent\textbf{The supervision target matters independently.}
We next hold the full feature set and architecture fixed and vary only the
training target. Training on a CER proxy reduces NonRelPrune from
$0.212$ to $0.130$---a $39\%$ relative reduction. Because the features,
architecture, and training procedure are otherwise unchanged, this drop
is attributable to the training target alone. This model still exceeds
BERT (CER, Gemma3) in NonRelPrune ($0.130$ vs.\ $0.113$), showing that
discriminative features retain some value even under the wrong
objective---but neither factor alone is sufficient.

We also examine whether selector scores correlate with discriminative
utility by partitioning entities into ten deciles by selector score and
computing mean OER log-odds per decile. BERT (CER, Gemma3) is nearly
flat ($0.10 \to 0.61$); the CER-trained model with OER features is
intermediate ($0.22 \to 0.92$); the full OER-proxy model rises steeply
($0.20 \to 1.28$--$1.30$). Candidate-set support features drive most of
the gain, but OER supervision is required to realize it fully.

\smallskip
\noindent\textbf{Mechanistic summary.}
The gains do not come from a novel feature family alone. Candidate-set
support provides the main signal, and the supervision target matters
independently: switching from OER to CER reduces NonRelPrune by $39\%$.
Observable distributional features become effective when entity selection
is aligned with OER rather than CER. We discuss the structural origin of
this mismatch in Section~\ref{sec:discussion}.

\section{Discussion}
\label{sec:discussion}

\noindent\textbf{Why the field got here.}
The CER--OER gap does not come from careless annotation or weak models. It arises because standard ad hoc collections provide document-level relevance judgments, but not entity-level discriminativeness labels. Entity supervision is therefore usually derived from document qrels, which produces CER-like or sparsity-biased labels and gives little signal about whether an entity separates relevant from non-relevant documents. Thus, the CER--OER gap is largely invisible during training.

A companion study~\cite{chatterjee2026entitiesretrievalsignalssystematic} shows that this problem is structural: binary derivation discards $83.6\%$ of discriminative entities by rewarding rarity rather than utility, and creates a $51$:$1$ negative-to-positive label imbalance. Section~\ref{sec:coverage_ceiling} shows the consequence: under current benchmarks, coverage and discrimination are difficult to optimize together, and post-hoc OER filtering hurts ranking by reducing relevant-document coverage. Closed-world evaluation further reinforced this proxy by hiding the coverage cost of discriminative entities and making broad, topically plausible entities appear more useful than they are in open-world retrieval.

\smallskip
\noindent\textbf{Implications for annotation practice.}
Entity annotation should move beyond asking whether an entity is topically related to a query. The more useful question is whether observing that entity increases the odds of relevance within the candidate set and linking environment. Future benchmarks should therefore expose annotators to corpus evidence, such as entity frequencies in relevant and non-relevant retrieved documents, rather than asking for topical judgments in isolation. Collections such as CODEC~\cite{mackie2022codec} are a step forward, but entity benchmarks should target observable retrieval utility more directly.

\smallskip
\noindent\textbf{Implications for system design.}
Entity-aware retrieval should treat candidate-set discriminative statistics as first-class signals. Local frequency, rank-weighted mention counts, and local IDF are available from first-stage retrieval output and require no new annotation. Our results suggest that the bottleneck is not representation quality alone, but the supervision target: semantic entity models still fail when optimized toward CER, while simple OER-aligned models produce stronger open-world gains. The same issue may arise in entity-based RAG filtering, where entities are often selected for topical accuracy rather than for their ability to discriminate relevant from non-relevant evidence.

\section{Conclusion}
\label{sec:conclusion}

Entity-aware retrieval has long assumed that useful entities are those
semantically related to the query. We show that this assumption is
insufficient in a precise and consequential way. Conceptual and observable
entity relevance are weakly aligned: topically plausible entities are often
poor retrieval signals, while discriminative entities are often peripheral.
This gap is systematic across collections and annotation sources, including
human entity judgments.

CER-trained entity selectors systematically select weakly discriminative
entities. They provide little practical filtering benefit and yield little
open-world improvement, regardless of architecture or supervision source.
Across 193 unsupervised entity-selection configurations, no method
simultaneously achieves high relevant-document coverage and high
discriminative precision. The ceiling is set by the benchmark environment,
not only by model quality.

Aligning supervision with OER breaks this pattern. In our controlled
fixed-ranker setting, OER-aligned selection improves open-world MAP by
$0.051$ over BM25+RM3 and $0.033$ over the best CER-based selector. This gain substantially exceeds the variation among supervised
CER-based selectors. Non-relevant
pruning improves by up to $10\times$ across collections, confirming that the
gain is not Robust04-specific. The bottleneck is the supervision target.

The central contribution of this paper is the CER--OER distinction itself:
entity labels and entity signals are not the same thing. Progress in
entity-aware retrieval will require choosing entities not because they are
\emph{about} the query, but because their observed presence provides useful
evidence for ranking. It will also require annotation and evaluation
infrastructure that measures this observable utility directly.

\balance
\bibliographystyle{ACM-Reference-Format}
\bibliography{references}
\end{document}